# A Total Variation Denoising Method Based on Median Filter and Phase Consistency

Huang Shuo[1,2], Wan Suiren[1,*]

1. School of Biological Sciences and Medical Engineering, Southeast University, Nanjing, 210096, China.
2. Shanghai United-imaging Healthcare Co., Ltd, Jiading district, Shanghai, 201807, China.
*. Corresponding author: Wan: srwan@seu.edu.cn.

**Abstract**: The total variation method is widely used in image noise suppression. However, this method is easy to cause the loss of image details, and it is also sensitive to parameters such as iteration time. In this work, the total variation method has been modified using a diffusion rate adjuster based on the phase congruency and a fusion filter of median filter and phase consistency boundary, which is called "the MPC-TV method". Experimental results indicate that MPC-TV method is effective in noise suppression, especially for the removing of speckle noise, and it can also improve the robustness of iteration time of TV method on noise with different variance.

## 1. Introduction

It is well known that the image noise suppression is an essential research topic in the field of digital image processing, especially in medical image processing [1-8]. The noise can be classified into additive noise such as the Gaussian noise and the multiplicative noise such as speckle noise and salt-and-pepper noise, which is a special form of speckle noise [9]. Since J. S. Lee points out that the fully developed speckle noise can be approximately expressed by the same equation of the additive noise [10], the model of noised image with either Gaussian noise or speckle noise can be expressed using Equation (1):

$$u_{noised} = u_{clean} + n \qquad (1)$$

where $u_{noised}$ and $u_{clean}$ are the noised image and the desired clean image, respectively. $n$ is the additive noise.

In recent years, a lot of noise suppression methods have been proposed [11-12]. These approaches include the wavelet denoising approach [7, 13, 14], the Gaussian filter [15] and the contourlet transform method [16]. However, these approaches are easy to damage the boundary and texture. Another famous approach is the variational image de-noising method, which is based on the least-square estimation [17]. But it is an isotropic diffusion method and is also weak in the protection of boundary, texture and details since the diffusion term of this method is defined in the $L^2$-norm space.

In 1992, Rudin proposed a total variation (TV) method [9], which aims to minimize the total variation energy $E[u]$ of image $u$, as shown in Equation (2):

$$\min_u E[u] = \int_\Omega |\nabla u| dxdy + \frac{\lambda}{2} \int_\Omega |(u - u_0)^2| dxdy \qquad (2)$$

where $\Omega$ represents the processed area and $\lambda$ denotes the global scale factor. With the help of the Euler-Lagrange equation, we have:

$$\nabla \cdot \frac{\nabla u}{|\nabla u|} - \lambda(u - u_0) = 0. \qquad (3)$$

The TV method has strong ability in both noise elimination and boundary protection, since the diffusion coefficient $1/|\nabla u|$ is related to the gradient of the image. It can both prevent the boundaries and textures and remove noise in the smooth regions of the image, since the diffusion will stop at regions with large gradient.

However, this method has some drawbacks. Firstly, when it comes to the image with severe noise, it may regard noise as boundary and cause the staircase effect. Secondly, although the TV method can prevent the textures and boundaries to some extent, it can still cause the loss of information in these regions [17-19], especially under a large iteration time. Therefore, choosing proper parameters, especially the iteration time, is essential to acquire good noise removal performance. However, since it is difficult to estimate the variance of noise, the image is easily to be over-smoothed.

## 2. Method

In this work, the MPC-TV method based on the total variation method has been put forward. In this method, the TV method has been modified using an adjust factor based on the phase congruency to further adjust the diffusion coefficient, and a fusion filter of median filter and phase consistency boundary to better remove noise.

2.1 *Solving the total variation equation by partial differential equation*

The Equation (3) can be solved by Equations (4) using partial differential equation (PDE) under given time step $dt$, global scale factor $\lambda$, iteration time $N$, and a gradient regularization parameter $\varepsilon$. [20]

$$\frac{\partial u}{\partial t} = \frac{1}{|\nabla u|} \cdot u_{\xi\xi} + \lambda \times (u_0 - u) \tag{4}$$

where $u_{\xi\xi}$ is the second derivative of image $u$ in the direction of edges and textures, which is perpendicular to the direction of image gradient and can be acquired by Equations (5) and (6) using the first- and second-order derivative of image $u$.

$$\xi = \frac{\nabla^{\perp} u}{|\nabla u|} = \frac{(-u_y, u_x)}{\sqrt{u_x^2 + u_y^2}} \tag{5}$$

$$u_{\xi\xi} = \frac{u_y^2 u_{xx} - 2 u_x u_y u_{xy} + u_x^2 u_{yy}}{|\nabla u|^2} \tag{6}$$

Moreover, a gradient regularization parameter $\varepsilon$ was employed when calculate Equations (4) and (6), as shown in Equation (7) and (8):

$$|\nabla u| = \sqrt{u_x^2 + u_y^2 + \varepsilon^2} \tag{7}$$

$$u_{\xi\xi} = \frac{(u_y^2 + \varepsilon^2) u_{xx} - 2 u_x u_y u_{xy} + (u_x^2 + \varepsilon^2) u_{yy}}{|\nabla u|^2} \tag{8}$$

Table 1 shows the specific process of the TV method.

Table 1. The specific process of the TV method

| Algorithm 1. The TV method |
| --- |
| 1. Input: noised image $u$. |
| 2. Set parameters $dt, \lambda, \varepsilon$ and $N$. |
| 3. Let $u_0 = u$ |

```
4.   for  i = 1: 1: N
5.        Calculate the first- and second-order derivative of image  u_{i-1}
6.        Calculate  |∇u_{i-1}|  using Equation (7).
7.        Calculate  u_{i,ξξ}  using Equation (8).
8.        Obtain  u_{i-1}  by solving Equation (4)
9.   end
10.  Output: image  u_N.
```

## 2.2 The diffusion rate adjuster based on the phase congruency

In order to better protect the boundary, the maximum moment of phase congruency covariance (defined as $M$) has been employed to adjust both the diffusion coefficient and the global scale factor. This factor is an efficient edge detection operator, which is highly localized and has responses that are invariant to image contrast [21]. Equations (9)-(12) are the definition of this parameter.

$$M = \frac{1}{2}(c + a + \sqrt{b^2 + (a-c)^2} \tag{9}$$

where

$$a = \sum (PC(\theta)\cos(\theta))^2 \tag{10}$$
$$b = 2 \times \sum ((PC(\theta)\cos(\theta)) \cdot (PC(\theta)\sin(\theta))) \tag{11}$$
$$c = \sum (PC(\theta)\sin(\theta))^2 \tag{12}$$

where $PC(\theta)$ refers to the phase congruency value determined at orientation $\theta$, and the sum in Equations (10) - (12) is performed over the discrete set of orientations used (typically six).

The operator $M$ indicates the significance of the boundary, a larger $M$ represents a more significant boundary. In MPC-TV method, an adjust factor $g$ has been introduced to adjust the diffusion rate of the TV method based on the phase congruency. In the $i^{th}$ iteration of MPC-TV method, firstly, the maximum moment of phase congruency $M(u_{i-1})$ about the diffusion result of the $(i-1)^{th}$ iteration $u_{i-1}$ is obtained. Secondly, $M(u_{i-1})$ is normalized to $[0,1]$, which is defined as the normalized moment $M_{norm}(u_{i-1})$. Finally, the adjust factor can be obtained by Equation (13).

$$g_i = (1 - M_{norm}(u_{i-1}))^m \tag{13}$$

According to experiments in Section 3.1, we have $m = 2$.

With the help of the adjust factor, Equation (4) can be adjusted to Equation (14). In Equation (14), the diffusion rate on boundaries and textures is furtherly reduced, so that these details can be better prevented.

$$\frac{\partial u}{\partial t} = \frac{1}{|\nabla u|} \cdot g \cdot u_{ξξ} + \lambda \times g \cdot (u_0 - u) \tag{14}$$

Furthermore, the gradient regularization parameter $\varepsilon$ is changed to $\varepsilon_i = \varepsilon_0 \times g_i$ in our method.

## 2.3 The fusion filter of median filter and phase consistency boundary

In order to batter remove noise, another way is using other filters in the iteration process to further suppress the noise. Professor Gui and Liu employed the fuzzy-median in each iteration step of the anisotropic diffusion filter to remove the noise at edges and acquired good noise removal performance in Shepp-Logan head phantom [4, 22 - 24]. However, when facing images with complex textures and boundaries, the median filter needs improve, since it can severely damage details of images.

In this work, a fusion filter based on the median filter and the adjust factor in Equation (13) has been put forward, which is shown in Equation (15).

$$u_{med} = dt \times g_i \cdot Median\{u_{before\_med}\} + (1 - dt \times g_i) \cdot u_{before\_med} \tag{15}$$

where $Median\{u_{before\_med}\}$ represents the $3\ pixel \times 3\ pixel$ median filter, $u_{before\_med}$ is the input image to the fusion filter and $dt$ is the time step when solving Equation (14).

Table 2 shows the specific process of our proposed MPC-TV method.

Table 2. The specific process of MPC-TV method

| Algorithm 2. The MPC-TV method |
|---|
| 1. Input: noised image $u$. |
| 2. Set parameters $dt, \lambda, \varepsilon_0$ and $N$. |
| 3. Let $u_0 = u$ |
| 4. for $i = 1:1:N$ |
| 5.     Calculate the $M_{norm}(u_{i-1})$ using Equations (4a)-(4d). |
| 6.     Calculate $g_i$ using Equation (13). |
| 7.     Calculate $\varepsilon_i$ using $\varepsilon_i = \varepsilon_0 \times g_i$. |
| 8.     Calculate the first- and second-order derivative of image $u_{i-1}$ |
| 9.     Calculate $|\nabla u|$ using Equation (7). |
| 10.     Calculate $u_{\xi\xi}$ using Equation (8). |
| 11.     Obtain $u_{i\_before\_median}$ by solving Equation (14) |
| 12.     Calculate Equation (15) to get $u_i$ from $u_{i\_before\_median}$ |
| 13. end |
| 14. Output: image $u_N$. |

## 3. Results

In the experiments, the SNR (signal-to-noise ratio) was employed to quantitatively evaluate the noise suppress effect of MPC-TV method. It describes the similarity between the original clean image $u_{clean}$ and the de-noised image $u_N$, as shown in Equation (16):

$$SNR(u_N, u_{clean}) = 10 \times log_{10}\left(\frac{\sum_{j=1}^{M}(u_N(j)-\mu(u_N))^2}{\sum_{j=1}^{M}(u_N(j)-u_{clean}(j))^2}\right) \tag{16}$$

where $M$ is the number of pixels on each image, and $\mu(u_N)$ denotes the average of image $u_N$. The Lena image under zero-mean Gaussian noise with different noise variance (defined as $\sigma_n^2$) is employed in our experiment to verify the noise-suppression effect of MPC-TV method.

3.1 *Obtaining the proper power $m$ in the adjust factor*

The power $m$ in Equation (13) can greatly influence the noise remove effect of MPC-TV method. In this work, different values from 1 to 5 was tested in Lena image with zero-mean Gaussian noise under variance $\sigma_n^2 = 300$. In our experiments, we have $\varepsilon = 0.02$, $\lambda = 0.14$, and $dt = \varepsilon/5 = 0.004$.

Figure 1 shows the SNR of image $u_N$ with $N$ varies from 1 to 35. Moreover, to make a better comparison, the highest SNR and the beat iteration time under each power are listed in Table 3.

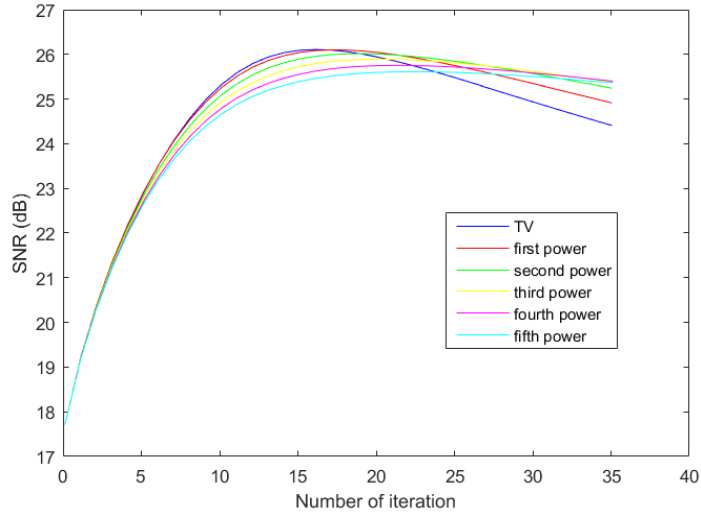

Figure 1. The SNR of results under different iteration number and different power in the adjust factor.

Table 3. The best iteration time and the highest SNR under different power in the adjust factor

| m | Original TV | 1 | 2 | 3 | 4 | 5 |
|---|---|---|---|---|---|---|
| Best iteration time | 16 | 17 | 19 | 20 | 21 | 22 |
| Highest SNR (dB) | 26.11 | 26.10 | 26.01 | 25.89 | 25.75 | 25.61 |

According to Figure 1 and Table 3, although the original TV Method has the largest value of the highest SNR and requires least iteration time to reach its highest SNR, when the iteration time is larger than the best number, the SNR decreases rapidly. With the increase of the power, this phenomenon is improved. However, at the same time, the best iteration time increases, and the highest SNR decreases. Therefore, the $2^{nd}$ power is selected, since it can significantly improve the decrease on SNR of original TV method under large iteration times, and with only slight decrease on the highest SNR and short increase on the best iteration time.

3.2 *The robustness of parameters about MPC-TV method*

The robustness of parameters about different noise variance ($\sigma^2 = 100, 200, 300, 400\ and\ 500$) and iteration times (from 1 to 35) of the original TV method and MPC-TV method are tested using Lena image under zero-mean Gaussian image, and results are shown in Figure 2. In our test, we have $\lambda = 0.14,\ \varepsilon = 0.02,\ dt = \varepsilon/5 = 0.004$.

Figure 2 shows that for the original TV method, these given parameters are only suitable for $\sigma^2 = 500$ among the 5 given conditions. It is not easy to find a proper iteration time that can fit all the five conditions. However, for MPC-TV method, the decent part of all the 5 curves in Figure 2(b) are flatter than that of the 5 curves in Figure 2(a). Although we still cannot find a proper iteration time to suit all the five conditions, repeating the iteration for about 25-30 times can acquire acceptable results when $\sigma^2 = 300, 400\ and\ 500$. Therefore, MPC-TV method has higher robustness to the noise variance and iteration time than the TV method.

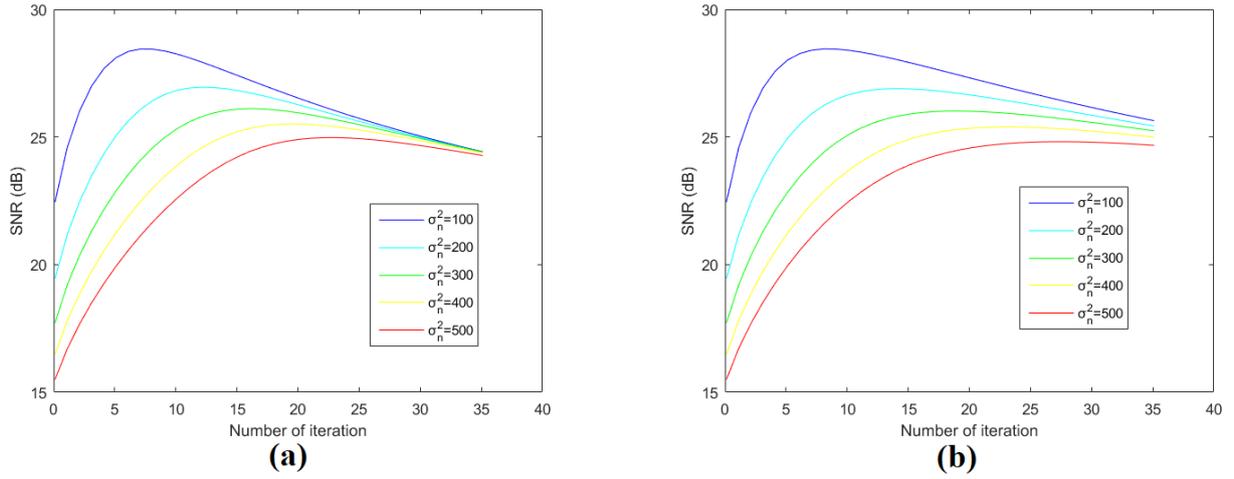

Figure 2. The robustness of parameters about different noise variance and iteration times of the original TV method and MPC-TV method. (a) shows results of the original TV method, and (b) shows results of the MPC-TV method.

## 4. Discussion

In this section, the boundary and texture protection effect of MPC-TV method is studied first, then the de-noising effects of MPC-TV method on speckle noise and salt-and-pepper noise have been analyzed.

### 4.1 *The boundary and texture protection effect of MPC-TV method*

The de-noising effect of TV method and MPC-TV method on Lena image under variance $\sigma_n^2 = 300$ has been employed to test the texture protection effect of the MPC-TV method. In our experiments, we also have $\varepsilon = 0.02$, $\lambda = 0.14$, and $dt = \varepsilon/5 = 0.004$.

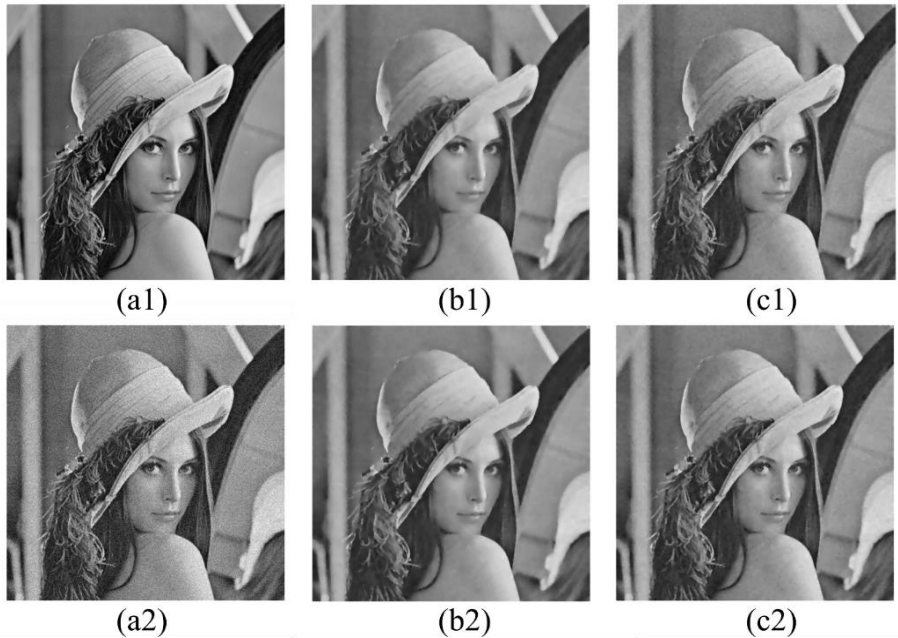

Figure 3. the test results of TV method and MPC-TV method when $\sigma^2 = 300$. (a1), (a2) are the clean and noised images of Lena, respectively. (b1) and (c1) are the de-noised images using TV method. (b2) and (c2) are the de-noised images using MPC-TV method. Iteration times of (b1), (b2), (c1) and (c2) are 26, 29, 16 and 19, respectively.

Figure 3 shows the test results of TV method and MPC-TV method when $\sigma^2 = 300$, and the SNR of each image in Figure 3 is listed in Table 4. Figures 4(c1) and 4(c2) are the result with highest SNR among all the tested iteration times of each method, and the corresponding iterative times are 16 and 19, respectively. Figures 4(b1) and 4(b2) are obtained using iteration times 26 and 29, which are ten times larger than the best iteration time of each method, respectively. Figures 4(c1) and 4(c2) have similar visual performance and similar SNR, however, compared with Figure 4(b1), the textures and boundaries are clearer in Figure 4(b1), which furtherly proves that the MPC-TV method is more robust on iteration time and also indicates that the MPC-TV method is efficient in texture and boundary protection.

Table 4. The SNR and MSSIM of images in Figure 3.

| Noised image | | Image de-noised by TV | | | Image de-noised by MPC-TV | | |
|---|---|---|---|---|---|---|---|
| SNR (dB) | MSSIM | Iteration time | SNR (dB) | MSSIM | Iteration time | SNR (dB) | MSSIM |
| 17.7152 | 0.3918 | 26 (Figure 3(b1)) | 25.3630 | 0.8388 | 29 (Figure 3(b2)) | 25.6228 | 0.8436 |
| | | 16 (Figure 3(c1)) | 26.1094 | 0.8474 | 19 (Figure 3(c2)) | 26.0157 | 0.8488 |

To further compare these results, mean structural similarity index measurement (MSSIM) [22, 25-29] and SSIM index map [25, 29] are employed, as shown in Equations (17) and (18), where $\sigma_{image}$ is the standard variation of image, $cov(u_N(j), u_{clean}(j))$ denotes the covariance of images in square windows centered by pixel $j$ in images $u_N$ and $u_{clean}$. $c_1$, $c_2$ and $c_3$ are three constant parameters set, and $c_1 = (0.01 \times max)^2$, $c_2 = (0.03 \times max)^2$, $c_3 = 1/2 \times c_2$, where max is the maximum possible pixel value of the image. In our experiments, the pixels are represented using 8 bits per sample, so max = 255.

$$SSIM(u_N(j), u_{clean}(j)) = \frac{2\mu(u_N(j))\mu(u_{clean}(j))+c_1}{\mu^2(u_N(j))+\mu^2(u_{clean}(j))+c_1} \cdot \frac{2\sigma_{image}(u_N(j))\sigma_{image}(u_{clean}(j))+c_2}{\sigma_{image}^2(u_N(j))+\sigma_{image}^2(u_{clean}(j))+c_2} \cdot \frac{cov(u_N(j),u_{clean}(j))+c_3}{\sigma_{image}(u_N(j))\sigma_{image}(u_{clean}(j))+c_3} \quad (17)$$

$$MSSIM(u_N, u_{clean}) = \frac{1}{M}\sum_{j=1}^{M}\left(SSIM(u_N(j), u_{clean}(j))\right)^2 \quad (18)$$

The SSIM index map is a matrix that has the same size with the image [25, 29]. To calculate the value of its elements, a local 8×8 square window which moves pixel-by-pixel over the entire image is used. At each step, the local statistics and SSIM index are calculated within the local window by Equation (17). And after the calculation of the SSIM index map, the mean value of the SSIM index map is calculated by Equation (18) to evaluate the overall image quality, which is the value of parameter MSSIM.

The MSSIM values of images in Figure 3 are listed in Table 4, and the SSIM index maps of Figures 3(b1), (b2), (c1) and (c2) are shown in Figure 4. Table 4 shows that Figures (c1) and (c2) have similar MSSIM value, however, after 10 more iterations, image de-noised by TV loses more information than image de-noised by MPC-TV method, since Figure 3(b1) has least MSSIM value among all four results.

As for the SSIM index map, differences between Figures 4(a1) and (a2) is shown in Figure 4(a3), and differences between Figures 4(b1) and (b2) is shown in Figure 4(b3). In Figures 4(a3) and (b3), the yellow and red regions denote the regions that the MPC-TV method has better performance, the green regions indicate regions where the MPC-TV method and the TV method have similar SSIM value, while in the blue regions, the TV method is better. According to Figure 4, although both the MPC-TV method and the TV method acquire best de-noising effect in Figures 3(c1) and (c2), in some regions of the textures and boundaries, the MPC-TV method can acquire better SSIM value. When it comes to the over-smoothed images in Figures 3(b1) and (b2), the texture and boundary protection effect of the MPC-TV method is much better than the TV method.

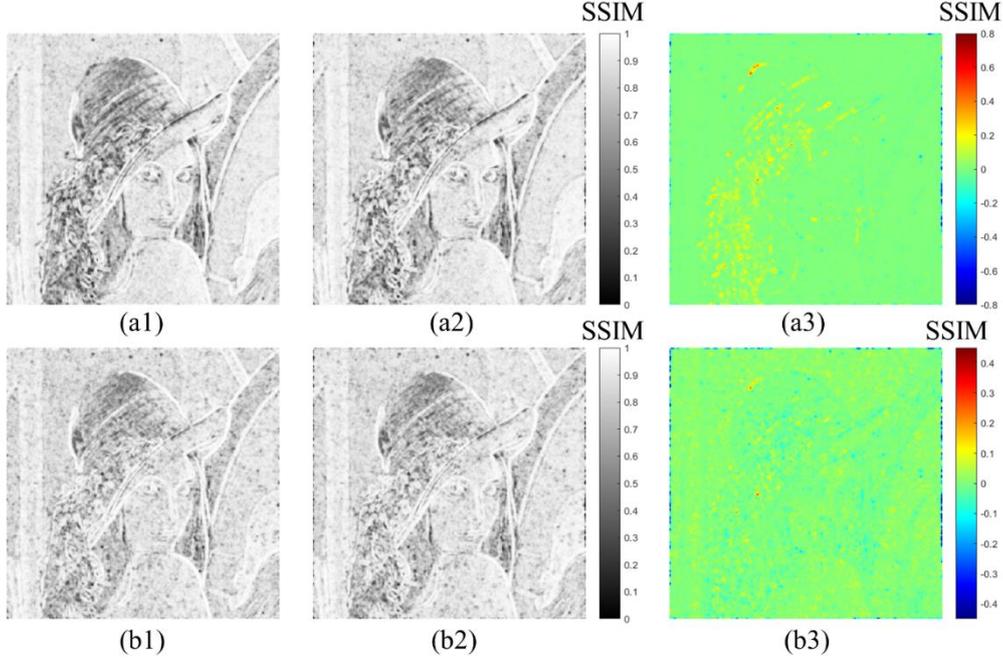

Figure 4. The SSIM index maps of Figures 3(b1), (b2), (c1) and (c2). (a1), (a2), (b1), (b2) are the SSIM index maps of Figures 3(b1), (b2), (c1) and (c2), respectively. (a3) is the map of differences between (a1) and (a2), (b3) is the map of differences between (b1) and (b2). In Figures (a3) and (b3), the yellow and red regions denote the regions that MPC-TV method has better performance, the green regions indicate regions where MPC-TV method and the TV method have similar SSIM value, while in the blue regions, the TV method is better.

4.2 *The de-noising performance of MPC-TV method on speckle noise*

The noise removal effects of MPC-TV method on Lena image with speckle noise is tested in this section. The variance of noise varies from $\sigma_n^2 = 500$ to $\sigma_n^2 = 900$, and the values of parameters are: $\varepsilon = 0.015$, $\lambda = 0.14$, and $dt = \varepsilon/5 = 0.003$.

Figure 5 shows the denoising performances of the TV method and the MPC-TV method with iteration times from 0 to 35. According to Figure 5, the highest SNR and SSIM value of MPC-TV method is higher than that of the TV method when the noise variance varies from $\sigma_n^2 = 500$ to $\sigma_n^2 = 900$. This indicates that MPC-TV method has better noise removal effect about speckle noise.

Moreover, on comparison with the de-noising of Gaussian noise, the curves of both SNR and MSSIM are flatter after reaching the highest value. Therefore, the MPC-TV method has higher robustness than the TV method on both iteration time and the variance of noise when removing speckle noise. This is because as multiplicative noise, unlike the Gaussian that distributes evenly in the image, the distribution of speckle noise in a certain region of image depends on the grayscale value of the clean image in that region and the intensity of texture of the clean image in that region. Usually the noise in the flat regions of images is more serious than the noise in texture regions. Therefore, reducing the diffusion coefficient by the adjust factor in the texture regions is more efficient to prevent the details of the image and improve the de-noising effect in the de-noising of speckle noise than in the de-noising of Gaussian noise.

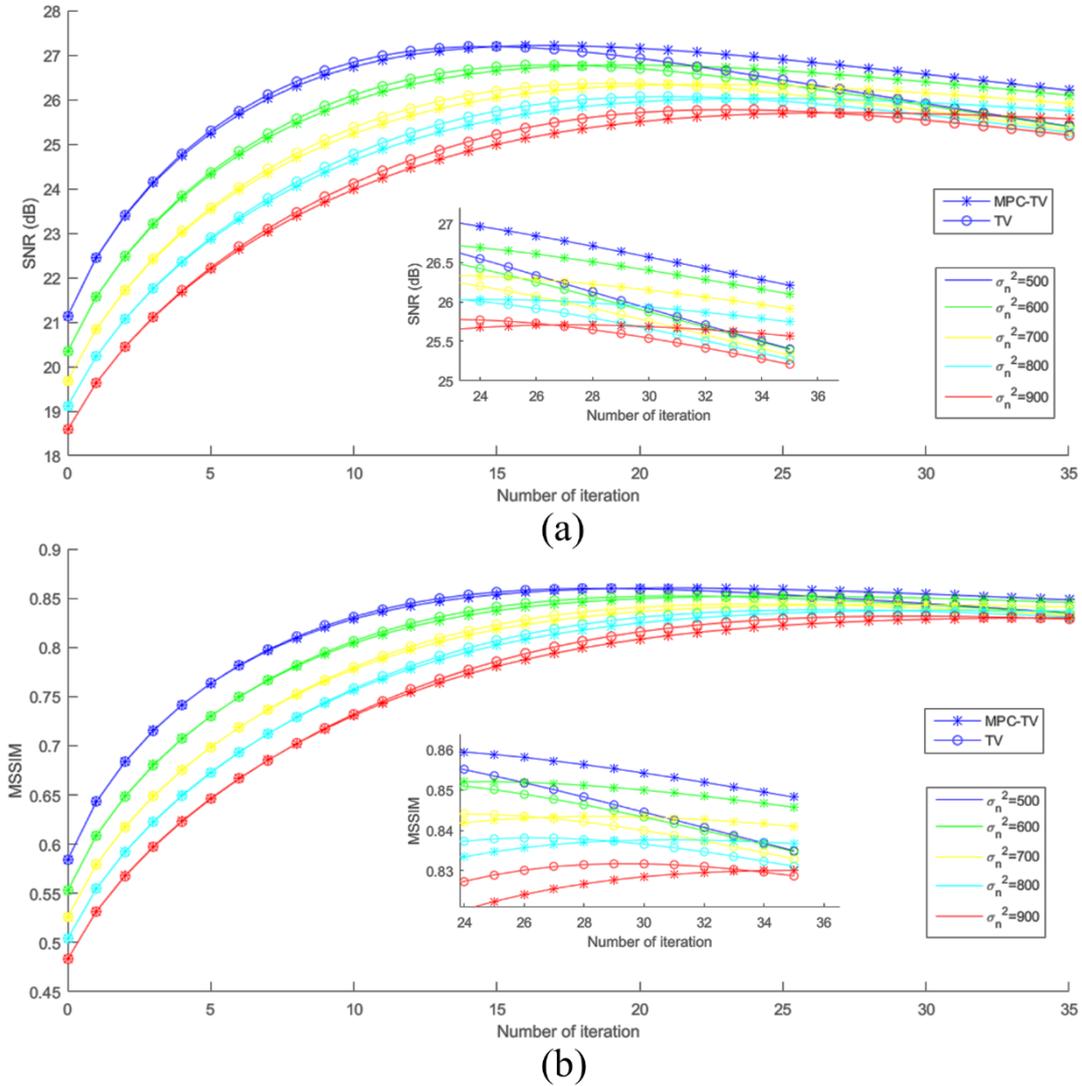

Figure 5. The de-noising performances of TV method and MPC-TV method of speckle noise on Lena image. (a) The SNR of image de-noised by different iteration times. (b) The MSSIM of image de-noised by different iteration times.

4.3 *The denoising effect of MPC-TV method on salt-and pepper noise*

Although MPC-TV method is efficient in the removal of Gaussian and speckle noise, its performance on salt-and-pepper noise still needs to be improved. Figure 6 shows the denoising performance of $(3\ pixel \times 3\ pixel)$ median filter, TV method and MPC-TV method on Lena image with salt-and-pepper noise. The variance of noise is $\sigma_n^2 = 900$, and the parameters are set to be: $\varepsilon = 0.03$, $\lambda = 0.14$, and $dt = \varepsilon/5 = 0.006$, respectively. The iteration time of TV method is 19, and that of the MPC-TV method is 25.

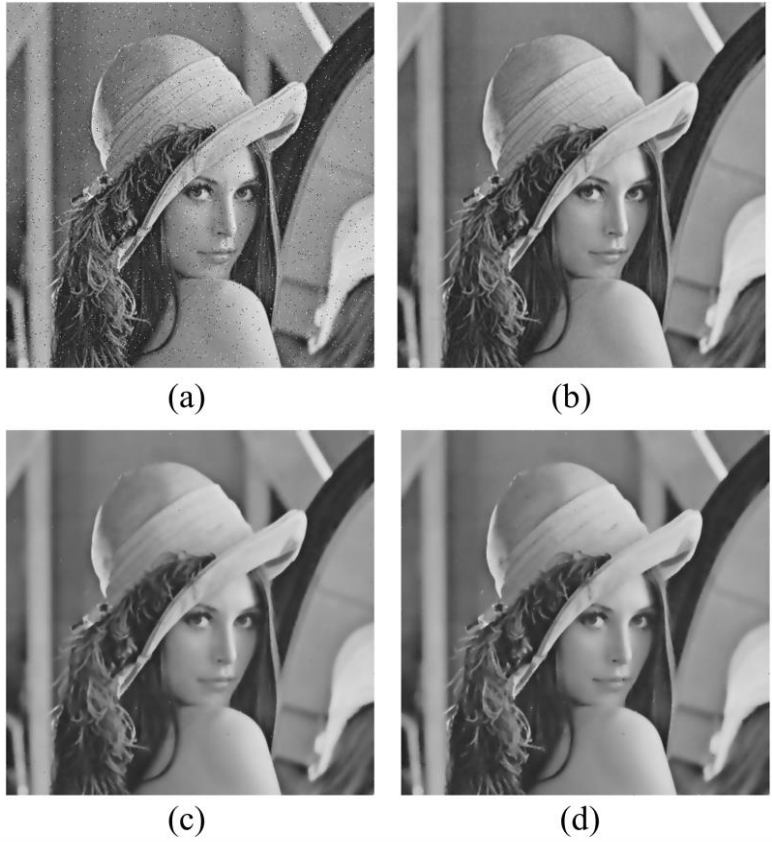

Figure 6. The de-noising performances of median filter, TV method and MPC-TV method of salt-and-pepper noise on Lena image. (a) is the noised image, (b) – (d) are results de-noised by the median filter, TV method and the MPC-TV method, respectively.

Table 5. The SNR and MSSIM of results in Figure 6

| Method | Noised image | Median filter | TV method | MPC-TV method |
|---|---|---|---|---|
| SNR (dB) | 18.4162 | 29.5868 | 25.1184 | 25.2238 |
| MSSIM | 0.6926 | 0.9214 | 0.8344 | 0.8356 |

Results in Figure 6 are quantitatively evaluated by parameters SNR and MSSIM, as shown in Table 5. Figure 6 and Table 5 shows that although the MPC-TV method has better performance in protecting the details of image than the TV method, the median filter has the best performance among all the three methods. Moreover, some noise points remained in the boundaries of the result of MPC-TV method, which also needs to be improved.

Moreover, we will also work on reducing the calculation time of MPC-TV method. Now, it takes about 0.6 second for a single iteration, which is about 17 times longer than the calculation time of one iteration for TV method. Therefore, reducing the calculation time is important in our future work.

## 5. Conclusion

In this study, a new total variation denoising method has been put forward, which is named after the MPC-TV method. In our algorithm, the maximum moment of phase congruency covariance has been employed to further adjust the diffusion rate, and a fusion filter of median filter and phase consistency boundary has also been put forward. Experimental results indicate that MPC-TV method is not only effective in noise suppression, especially for the removing of speckle noise, but also can improve the robustness of TV method on both iteration time and noise with different variance, since it can prevent the textures and boundaries of images.

Future work includes improving the de-noising effect of MPC-TV method on salt-and-pepper noise, and reducing the calculation time of the MPC-TV method.

**Acknowledgements:** The authors thank Professor He Siyuan, Zhou Ping, Sun Yu and Doctor Shi Hao from Southeast University for providing the experimenting tools for our study.

This work was supported by the National Key R&D Program of China under grant number 2017YFC0112801, the National Natural Science Foundation of China (NSFC) under grant numbers 61127002, 11572087, School of Biological Sciences and Medical Engineering of Southeast University under grant number 3207037434, the Project of Texas Instruments (TI) Company: Development of Modular OCT Imaging System based on TI Chips under grant number 8507030129 and the Key Project of Special Development Foundation of Shanghai Zhangjiang National Innovation Demonstration Zone under grant number 1701-JD-D1112-030.

**References**
[1] Wang Y, Shao Y, Zhang Q, et al. Noise Removal of Low-dose CT Images Using Modified Smooth Patch Ordering. *IEEE Access*. 5 (2017) 26092-26103.
[2] Liu Y, Castro M, Lederlin M, et al. Edge-preserving denoising for intra-operative cone beam CT in endovascular aneurysm repair. *Computerized Medical Imaging & Graphics*. 56 (2017) 49-59.
[3] Chen Y, Liu J, Xie L, et al. Discriminative Prior - Prior Image Constrained Compressed Sensing Reconstruction for Low-Dose CT Imaging. *Scientific Reports*. 7 (2017), 13868.
[4] Liu Y, Shangguan H, Zhang Q, et al. Median prior constrained TV algorithm for sparse view low-dose CT reconstruction. *Computers in Biology & Medicine*. 60(C) (2015) 117-131.
[5] Hasan M K, Shifat-E-Rabbi M, Lee S Y. Blind Deconvolution of Ultrasound Images Using $l_1$-Norm-Constrained Block-Based Damped Variable Step-Size Multichannel LMS Algorithm *IEEE Transactions on Ultrasonics, Ferroelectrics, and Frequency Control*. 63(8) (2016) 1116-1130.
[6] Zhao N, Basarab A, Kouamé D, et al. Joint Segmentation and Deconvolution of Ultrasound Images Using a Hierarchical Bayesian Model Based on Generalized Gaussian Priors. *IEEE Transactions on Image Processing*. 25(8) (2016) 3736-3750.
[7] Kaur S, Kaur S. MRI denoising using non-local PCA with DWT. in *Fourth International Conference on Parallel, Distributed and Grid Computing*. IEEE, (2017) 507-511.
[8] Phophalia A, Mitra S K. 3D MR image denoising using rough set and kernel PCA method. *Magnetic Resonance Imaging*. 36 (2017) 135-145.
[9] Rudin L I, Osher S, Fatemi E. Nonlinear total variation based noise removal algorithms. *Physical D Nonlinear Phenomena*. 60(1–4) (1992) 259-268.
[10] Lee J S. Digital image enhancement and noise filtering by use of local statistics. *IEEE Transactions on Pattern Analysis & Machine Intelligence*. PAMI-2(2) (1980) 165-168.
[11] Zhang Y, Wu L. Improved image filter based on SPCNN. *Science China Information Sciences F-Information Sciences*. 51(12) (2008) 2115-2125.
[12] Zhang Y, Dong Z, Phillips P, et al. Exponential Wavelet Iterative Shrinkage Thresholding Algorithm for compressed sensing magnetic resonance imaging. *Information Sciences*. 322(1) (2015)

115-132.

[13] Wan S, Raju B I, Srinivasan M A. Robust deconvolution of high-frequency ultrasound images using higher-order spectral analysis and wavelets. *IEEE Transactions on Ultrasonics, Ferroelectrics and Frequency Control*. 50(10) (2003) 1286-1295.

[14] Donoho D L. De-noising by soft-thresholding. *IEEE Transactions on Information Theory*. 41(3) (1995) 613-627.

[15] Seddik H, Tebbini S, Braiek E B. Smart Real Time Adaptive Gaussian Filter Supervised Neural Network for Efficient Gray Scale and RGB Image De-noising. *Intelligent Automation & Soft Computing*. 20(3) (2014) 347-364.

[16] Li D M, Zhang L J, Yang J H, et al. Research on wavelet-based contourlet transform algorithm for adaptive optics image denoising. *Optik*, 127(12) (2016) 5029-5034.

[17] Chan T F, Osher S, Shen J. The digital TV filter and nonlinear denoising. *IEEE Transactions on Image Processing*. 10(2) (2001) 231-241.

[18] Zhang H, Peng Q. Adaptive image denoising model based on total variation. *Opto-electronic Engineering*. 33(3) (2006) 50-53. (*in Chinese*)

[19] Niu H, Du Q, Zhang J. An Algorithm of Adaptive Total Variation Image Denoising. *Pattern Recognition and Artificial Intelligence*. 24(6) (2011) 798-803. (*in Chinese*)

[20] Wang Y. A new numerical realization algorithm based on the total variation model. *Journal of Yunnan University of Nationalities (Natural Sciences Edition)*. 19(3) (2010) 212-215. (*in Chinese*)

[21] Kovesi P. Phase Congruency Detects Corners and Edges. in *DICTA*. (2003) 309-318.

[22] Gui Z G, Liu Y. Noise reduction for low-dose X-ray computed tomography with fuzzy filter. *Optik*. 123(13) (2012) 1207-1211.

[23] Gui Z, Liu Y, He J. PML Algorithm for Positron Emission Tomography Combined With Nonlocal Fuzzy Anisotropic Diffusion Filtering. *IEEE Transactions on Nuclear Science*. 59(5) (2012) 1984-1989.

[24] Ling J, Bovik A C. Smoothing low-SNR molecular images via anisotropic median-diffusion. *IEEE Transactions on Medical Imaging*. 21(4) (2002) 377-384.

[25] Liu J, Hu Y, Yang J, et al. 3D Feature Constrained Reconstruction for Low Dose CT Imaging. *IEEE Transactions on Circuits and Systems for Video Technology*. 28(5) (2016) 1232-1247.

[26] Huang S, Zhou P, Shi H, et al. Image speckle noise denoising by a multi-layer fusion enhancement method based on block matching and 3D filtering. *The Imaging Science Journal*. 67(4) (2019) 224-235.

[27] Liu Y, Gui Z and Zhang Q. Noise reduction for low-dose X-ray CT based on fuzzy logical in stationary wavelet domain. *Optik*. 124(18) (2013) 3348-3352.

[28] Wang Z, Simoncelli E P, and Bovik A C. Multi-scale structural similarity for image quality assessment. in Asilomar conference on signals, systems and computers, (2003) 1398-1402.

[29] Wang Z, Bovik A C, Sheikh H R, et al. Image quality assessment: from error visibility to structural similarity. *IEEE Transactions on Image Processing*. 13(4) (2004) 600-612.